\documentclass[twocolumn,10pt,english,aps,prl,longbibliography]{revtex4-2}
\usepackage[T1]{fontenc}
\usepackage[utf8]{inputenc}
\usepackage{color}
\usepackage{babel}
\usepackage{verbatim}
\usepackage{titlesec}
\usepackage[dvipsnames]{xcolor}
\usepackage{amsbsy}
\usepackage{amsmath}
\usepackage{amstext}
\usepackage{graphicx}
\usepackage{afterpage}
\usepackage{placeins}
\usepackage{enumerate}
\usepackage{multirow}
\usepackage{amssymb}
\usepackage{units}
\usepackage[unicode=true,pdfusetitle,
bookmarks=true,bookmarksnumbered=false,bookmarksopen=false,
 breaklinks=false,pdfborder={0 0 0},pdfborderstyle={},backref=false,colorlinks=true]{hyperref}
\hypersetup{urlcolor=blue,citecolor=blue}

\newcommand{\tsp}{\hspace{0.4mm}}

\newcommand{\noop}[1]{}

\global\long\def\phfrac{\vphantom{\dfrac{x}{x}}}%
\global\long\def\img{\tsp\textrm{i}\tsp}
\global\long\def\ccdot{\tsp\cdot\tsp}%
\global\long\def\plus{\tsp+\tsp}%



\begin{document}

\title{Disorder-induced damping of spin excitations in Cr-doped BaFe$_2$As$_2$}

\author{Marli R. Cantarino$^{1\dag}$,  Rafael M. P. Teixeira$^{2}$, K. R. Pakuszewski$^{3}$, Wagner R. da Silva Neto$^{4}$, Juliana G. de Abrantes$^{5}$, Mirian Garcia-Fernandez$^{6}$, P. G. Pagliuso$^{7}$, C. Adriano$^{3}$, Claude Monney$^{8}$, Thorsten Schmitt$^{9}$, Eric C. Andrade$^{2}$, and Fernando A. Garcia$^{2\dag}$}

\affiliation{$^{1}$European Synchrotron Radiation Facility, BP 220, F-38043 Grenoble Cedex, France}
\affiliation{$^{2}$Instituto de Física, Universidade de São Paulo, 05508-090 São Paulo, SP, Brazil}
\affiliation{$^{3}$Institut Quantique and Département de Physique, Université de Sherbrooke, 2500 boulevard de l’Université, Sherbrooke, Québec J1K 2R1, Canada}
\affiliation{$^{4}$Fakultät für Chemie und Mineralogie, Universität Leipzig, 04103 Leipzig, Germany}
\affiliation{$^{5}$School of Chemistry and Chemical Engineering, University of Surrey, Guildford GU2 7XH, United Kingdom}
\affiliation{$^{6}$Diamond Light Source, Harwell Campus, Didcot OX11 0DE, United Kingdom}
\affiliation{$^{7}$Instituto de Física “Gleb Wataghin”, UNICAMP, 13083-859, Campinas-SP, Brazil}
\affiliation{$^{8}$D\'epartement de Physique and Fribourg Center for Nanomaterials, Universit\'e de Fribourg, CH-1700 Fribourg, Switzerland}
\affiliation{$^{9}$Photon Science Division, Paul Scherrer Institut, 5232, Villigen PSI, Switzerland}

\affiliation{$^{\dag}$Corresponding authors M.R.C. (marli.cantarino@esrf.fr) and F.A.G. (fgarcia@if.usp.br)}

\begin{abstract}

In doped Hund’s metals, such as the iron-based superconductors, effects like charge doping and chemical pressure are often considered the dominant factors. Partial chemical substitution, however, inevitably introduces disorder. Here, we investigate spin excitations in Ba(Fe$_{1-x}$Cr$_x$)$_2$As$_{2}$ (CrBFA) by high-resolution resonant inelastic x-ray scattering (RIXS) for samples with $x = 0, 0.035,$ and $ 0.085$. In CrBFA, Cr acts as a hole dopant, but also introduces localized spins that compete with Fe-derived magnetic excitations. We found that the Fe-derived magnetic excitations are softened and damped, becoming overdamped for $x = 0.085$. At this doping level, complementary angle-resolved photoemission spectroscopy measurements (ARPES) show increased electronic localization and a suppression of the nematic $d_{xz}/d_{yz}$ band splitting present in the parent compound. We thus propose a localized spin model that explicitly incorporates substitutional disorder and Cr local moments, successfully reproducing our key observations. Our findings reveal a case where disorder dominates over charge doping in the case of a Hund’s metal. 

\end{abstract}
\maketitle


\textit{Introduction:} Electronic correlated materials often display multiple competing interactions, each associated with one or more energy scales. Partial chemical substitution tunes interaction energy scales to stabilize a desired ground state. This gives rise to rich composition ($x$) vs. temperature ($T$) phase diagrams, characteristic of correlated systems \citep{canfield_new_2020}. 

In iron-based superconductors (FeSCs) \citep{fernandes_iron_2022}, chemical substitutions often induce the suppression of the antiferromagnetic (AFM) ordered phase and the emergence of high-temperature superconductivity (HTSC) \citep{ni_phase_2009,sefat_superconductivity_2008,li_superconductivity_2009,chu_determination_2009,jiang_superconductivity_2009,saha_superconductivity_2010,rotter_superconductivity_2008}. While these effects are typically attributed to charge doping or chemical pressure \cite{canfield_feas-based_2010}, the role of disorder, an inevitable consequence of substitution, remains unclear. Early scenarios for the physics of FeSCs proposed that disorder could be the driving mechanism behind the observed $x$ vs. $T$ phase diagrams \citep{wadati_where_2010,berlijn_transition-metal_2012} but subsequent approaches favored charge doping and chemical pressure effects \cite{fernandes_low-energy_2016,chubukov_pairing_2012}.

The parent compound BaFe$_2$As$_2$ exhibits a stripe-type AFM transition at $T_{\text{N}} \approx 134$ K, accompanied by a tetragonal-to-orthorhombic structural transition \citep{rotter_spin-density-wave_2008,kim_character_2011}. Substitutions suppressing these transitions may or may not induce superconductivity (SC). For instance, Mn substitution (MnBFA) introduces minimal charge doping, mostly affecting the electron pockets \citep{cantarino_incoherent_2023,suzuki_absence_2013,texier_mn_2012}, and some attention was devoted to the effect of disorder acting as the main tuning knob in this phase diagram \citep{fernandes_suppression_2013,gastiasoro_enhancement_2014, inosov_possible_2013,tucker_competition_2012}. In contrast, Cr substitution (CrBFA) acts as a hole dopant \citep{2024_cantarino_CrBFA}, but the evolution of $T_{\text{N}}$ in CrBFA and MnBFA follows a similar trend as a function of $x$, with no emergence of SC for both substitutions. Therefore, charge doping alone cannot explain their phase diagrams.

Previous studies have shown that hole doping enhances electronic correlations in BaFe$_2$As$_2$ \citep{hardy_evidence_2013,lafuerza_evidence_2017}, an effect often attributed to proximity to a Mott insulating state. However, experimental observations show that the ground state remains metallic even in regimes with localized magnetism. This apparent contradiction can be resolved by the concept of spin-orbital separation in multiorbital systems: the Hund’s coupling delays spin screening relative to orbital screening, leading to the so-called Hund’s metal behavior. BaFe$_2$As$_2$ is currently recognized as a Hund’s metal, where strong Hund’s coupling and orbital selectivity govern the degree of electronic correlations, which are expected to increase further with hole doping toward half-filling \citep{haule_coherenceincoherence_2009,Hardy_Hund_holedoping,2017_pelliciari_Co-KBFA,georges_hund-metal_2024}. While K substitution on the Ba site preserves itinerancy and superconductivity even at high hole doping, direct substitution at the Fe site, as in the case of Cr, introduces carriers that enhance localization and Fe-sublattice disorder. Resonant inelastic x-ray scattering (RIXS) is proven to be a powerful tool to investigate magnetism with the contrasting itinerant character of electrons and localized character of spins within FeSC phase diagrams \citep{pelliciari_intralayer_2016, 2017_pelliciari_Co-KBFA, lu_spin-excitation_2022,liu_nematic_2024}.

In this work, we investigate how spin excitations originating from Cr and Fe compete for the ground state properties across the $x$ versus $T$ phase diagram of CrBFA. To this end, we performed high-resolution RIXS, complemented by angle-resolved photoemission spectroscopy (ARPES) experiments and theoretical modeling. Our combined experimental and theoretical analysis shows that the key features of the data can be understood within a localized spin model, where disorder effects introduced by Cr moments play a central role. This reveals a case of a correlated metal in which disorder dominates over the charge doping effect. 

Our results explain why the suppression of the AFM order in CrBFA and MnBFA does not follow the expected trend for charge doping, where Cr nominally introduces twice as many holes as Mn \citep{thaler_physical_2011,cantarino_incoherent_2023}. They may also help explain the $x$–$T$ phase diagrams of Cr/Mn substituted CaKFe$_4$As$_4$, where the suppression of both HTSC and AFM order similarly deviates from charge doping expectations \citep{xu_superconductivity_2022,xu_superconductivity_2023}. Altogether, our findings suggest that the absence of SC in these systems is likely linked to the joint effect of increasing correlation and disorder effects on the Fe lattice introduced by Cr and Mn impurities.


\textit{Materials and methods:} Ba(Fe$_{1-x}$Cr$_{x}$)$_{2}$As$_{2}$ single crystals, with different concentrations of Cr, were grown using the In-flux method \citep{garitezi_synthesis_2013}. Sample characterization methods are described in the Supplemental Material \ref{appB_exp} \citep{sefat_absence_2009,marty_competing_2011,clancy_high-resolution_2012}.

The $x = 0, 0.035,$ and $ 0.085$ samples (denominated as BFA, Cr$3.5\%$, and Cr$8.5\%$, respectively) were selected for RIXS experiments at the I21 beamline at Diamond Light Source \citep{2022_Zhou_I21_DLS_RIXS}, with a detected photon energy resolution of $\approx 30$ meV. The energy of the absorption spectra $L_3$ edge maximum at $709.5$ eV was chosen for the momentum-dependent studies. The methods of the sample preparation and data acquisition are detailed in the Supplemental Material \ref{appB_exp}.

\begin{figure*}
    \centering
    \includegraphics[width=\textwidth]{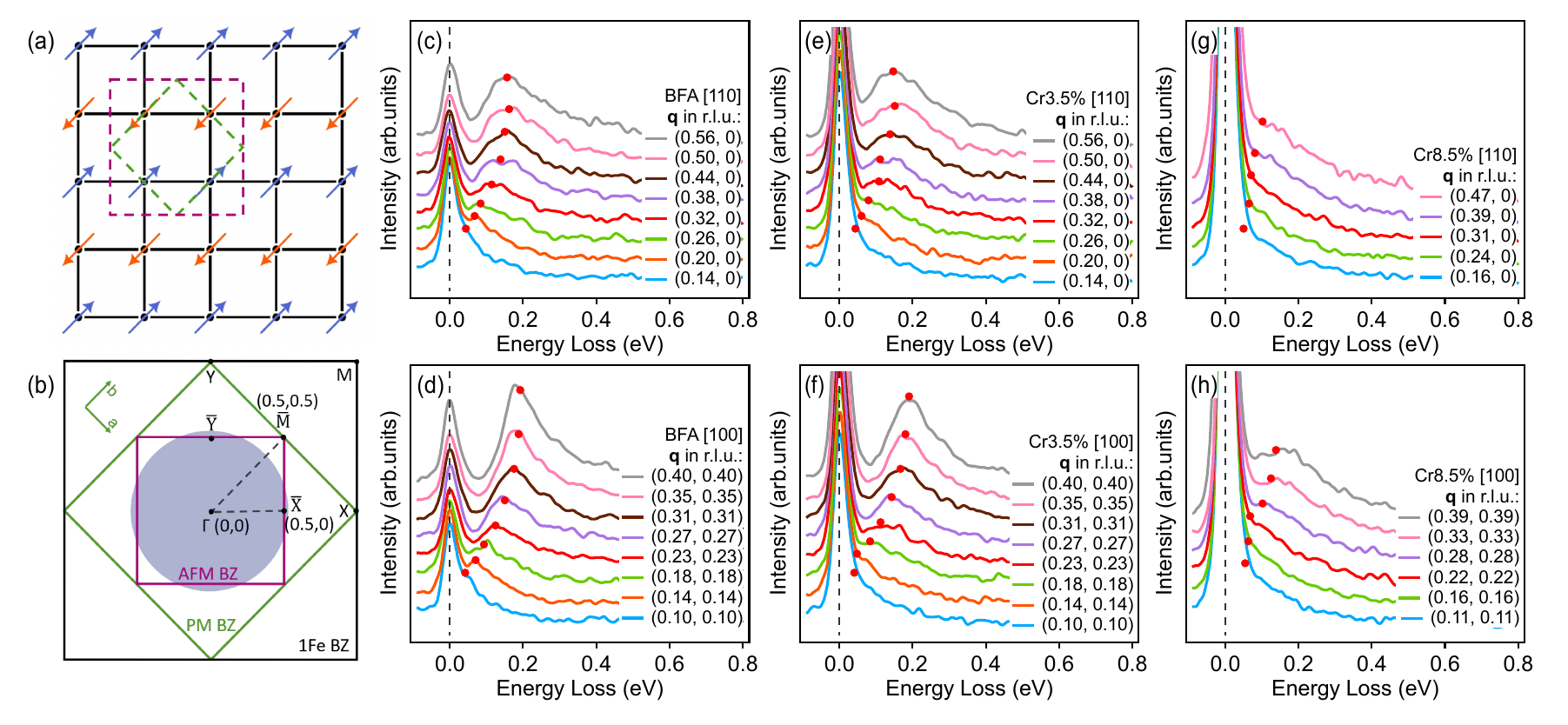}
    \caption{Overview of RIXS results. (a) 2D magnetic lattice with the stripe order. The green and magenta dashed squares show the PM and the AFM phases' unit cells. (b) The respective reciprocal spaces for the 2D first Brillouin zones, the circle indicates the probed $\mathbf{q}$-range, with the dashed lines showing the high-symmetry directions for the RIXS experiments. The 1Fe BZ (black square) is used for the theory. (c)-(h) Momentum-dependent RIXS spectra for the samples with $x = 0, 0.035$ and $0.085$ (BFA, Cr$3.5\%$, and Cr$8.5\%$) and for different directions after background subtraction. The data are shifted for better comparison. \label{fig:overview}}
\end{figure*}

We define the wavevector $\mathbf{q}$ in reciprocal lattice units (r.l.u.) as $(H,K) = (q_x,q_y)a/2\pi$, where $a = 5.598$, 5.590, and 5.616 {\AA} are the in-plane lattice parameters for BFA, Cr$3.5\%$, and Cr$8.5\%$, respectively. These correspond to the magnetic unit cell shown as a magenta dashed square in Fig. \ref{fig:overview}(a). Since the samples are twinned, we adopt $a = b$ and do not distinguish orthorhombic $a$ and $b$ directions. In this Brillouin Zone (BZ), we probe a maximum momentum of $|\mathbf{q}| \approx 0.56$ r.l.u., indicated by the blue circle in Fig. \ref{fig:overview}(b). High-symmetry points are labeled using the 1Fe BZ (black square) and consistently used for ARPES and theoretical results. ARPES acquisition details are as described in refs. \citep{cantarino_incoherent_2023,2024_cantarino_CrBFA} and in the Supplemental Material \ref{appB_exp}.

To model the magnetic excitations, we employ a minimal antiferromagnetic (AFM) XXZ $J_1-J_2$ model on a square lattice, with anisotropy favoring the spins to lie on the $S_x-S_y$ plane. Classically, this system exhibits checkerboard (Néel) order for $J_2<J_1/2$ and stripe order for $J_2>J_1/2$. Cr doping ($x$) is treated as a local disorder in $J_2$, the coupling controlling the transition between the ordered phases. At $x=0$, the system has stripe order; at $x=1$, it displays Néel order. For intermediate doping, we randomly replace a fraction $x$ of Fe sites with Cr, assigning $J_2$ between sites $i$ and $j$ as follows: (i) Fe--Fe, $J_{2,ij}=J_{2}(x=0)$; (ii) Cr--Cr, $J_{2,ij}=J_{2}(x=1)$; (iii) Fe--Cr, $J_{2,ij}=[J_{2}(x=1) + J_{2}(x=0)]/2$. The classical ground state is determined via Monte Carlo simulations and annealing. Excitation spectra are computed by integrating the semiclassical equations of motion: $d\vec{S}_{i}/dt=\vec{h}_i \times \vec{S}_i$, where $h_{i}^{\alpha}=\sum_{j}J_{ij}^{\alpha}S_j^{\alpha}$, is the local exchange field at site $i$ and $\alpha=x,y,z$. Finally, we Fourier transform $\vec{S}_i(t)$ to obtain the dynamical spin structure factor for comparison with experiments. Details are provided in the Supplemental Material \ref{appA_theory} \cite{Joyce1967,Manousakis1991,Auerbach1994,Parkinson2010,Ewings2008,Moessner1998,Conlon2009,Moessner2014,Shannon2018,Knolle2022,Miyatake_1986,Creutz1987,Dormand1980,Garanin2021,Frigo2005,Paddison2019,Smith2022}.  


\textit{Results and discussion:} In Fig. \ref{fig:overview}(c)-(h), we show an overview of our RIXS momentum-dependent spectra for each sample. The probed reciprocal space high-symmetry directions and range are represented by the black dashed lines and blue circle in Fig. \ref{fig:overview}(b). A background subtraction was performed to better visualize and fit the quasi-elastic region (which also features low-energy phonon excitations) and magnon excitations. The subtraction method is explained in the Supplemental Material \ref{appRIXS_fit} \cite{2009_Yang_weakcor,2010_Hancock_FeTe,zhou_persistent_2013}.

\begin{figure}
    \centering
    \includegraphics[width=\linewidth]{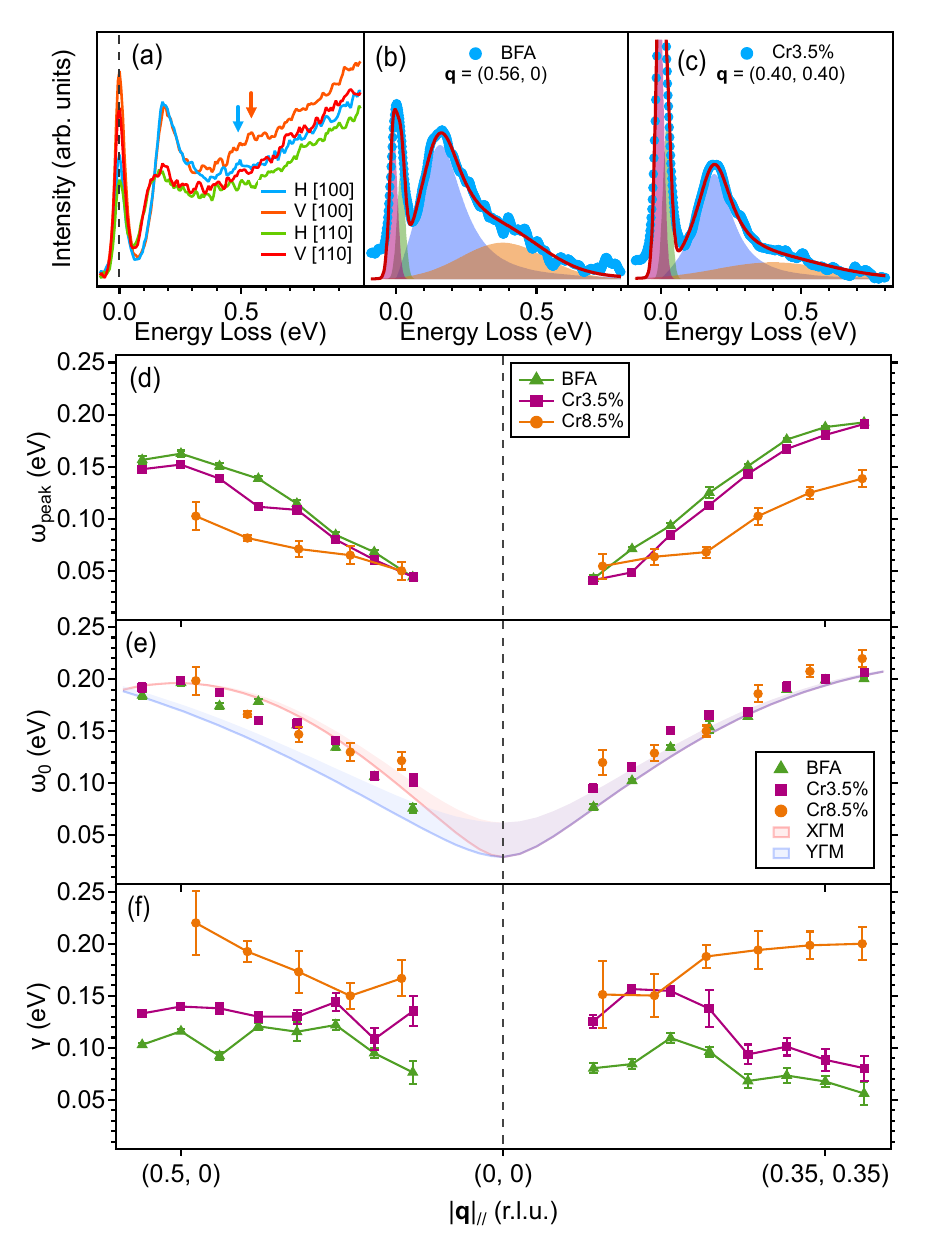}
    \caption{(a) Polarization-dependent RIXS spectrum of the parent compound (BFA) measured at $|\mathbf{q}| = 0.56$ r.l.u.. (b,c) Representative fits of the quasi-elastic and magnon contributions at maximum momentum transfer for the BFA and Cr$3.5\%$  samples, respectively. (d) Peak energy of the magnetic excitation $\omega_\text{peak}$, (e) bare frequency $\omega_0$, and (f) damping coefficient $\gamma$ as functions of momentum, for different samples and high-symmetry directions. In panel (e), the solid line shows the linear spin wave dispersion at $L = 0$, using parameters from inelastic neutron scattering fits \citep{harriger_nematic_2011}. The shaded area indicates the dispersion variation with $L$, considering the 3D Brillouin zone.\label{fig:dispersion}}
\end{figure}

The main magnon dispersion is clear even for the Cr$8.5\%$ substitution, for which the elastic line is enhanced and the magnon is softened. The magnon appears as a shoulder of the quasi-elastic peak at low $\mathbf{q}$ and is evident at maximum momentum transfer. To gain more quantitative insight, spectra were fitted to a model composed of an elastic (Gaussian), a phonon (Gaussian), a principal magnon peak, and a secondary inelastic magnonlike peak (Gaussian). 

The main magnon peak can be fitted to the imaginary part of the dynamic susceptibility of a damped harmonic oscillator (DHO) $\chi''(\omega)$, as shown in Equation \ref{eq:magpeak1} \cite{2016_lamsal_paramagnon,2018_Peng_magnon}, where the $\mathbf{q}$ dependency reveals the magnon dispersion, $I_0$ is an intensity constant, $\gamma$ is the damping coefficient, and $\omega_0$ is the undamped (bare) frequency of the spin excitation as a function of $\mathbf{q}$. The Bose-Einstein distribution is included to reproduce thermal effects:

\begin{equation}\label{eq:magpeak1}
    \chi''(\omega) =
    \frac{1}{1 - e^{-\hbar\omega / k_B T}}
    \cdot
    \frac{I_0 \gamma \omega}{(\omega^2 - \omega_0^2)^2 + 4(\gamma \omega)^2}
\end{equation}

We refer to the peak energy of the magnetic excitation as $\omega_\text{peak}$, i.e., the energy at $\chi''(\omega)$ reaches its maximum.

We note that it is necessary to include the second magnonlike peak in the fitting to accurately describe the shape and sharp width of the main magnon peak, especially for the highest $\mathbf{q}$ values and in the case of the parent compound. The inclusion of this peak is justified by the inspection of Fig. \ref{fig:dispersion}(a), wherein we show a direction and polarization survey of our results for the parent compound. As can be observed, a second excitation appears riding on top of the background fluorescence and peaking around $\approx450-550$ meV. It is clearer for the $[100]$ $(\pi,\pi)$ direction, as indicated by the orange and blue arrows.

Figs. \ref{fig:dispersion}(b) and (c) show examples of fittings for the BFA and Cr$3.5\%$ samples. The data after the background subtraction are shown as blue dots, the total fitting as the red line, and the shaded regions as the different fitting contributions, including the elastic line, phonon, main magnon, and secondary magnon-like peak, shown as pink, green, blue, and orange areas, respectively. All the fittings for different momentum points for all samples and directions are shown in the Supplemental Material \ref{appRIXS_fit}. The need for the secondary magnon peak is particularly evident along the $(\pi,\pi)$ direction, where the main magnon peak is sharp but notably asymmetric, which is not expected from the fitted model \citep{2018_Peng_magnon}. At lower $\mathbf{q}$ values and for the doped samples, the secondary magnon peak becomes less pronounced and may even vanish after background subtraction. The dispersive properties of this second peak can be observed for the $\mathbf{q}$ regions where its contribution has relevant weight to the fitting, close to the magnetic zone boundary, as shown in Fig. \ref{fig:2ndmag} of the Supplemental Material. This behavior supports the interpretation of the feature as a bimagnon, a bound state of two correlated spin flips on neighboring sites, with well-defined momentum and relatively long lifetime \citep{1995_theorybimagnon_PRB,2018_chaix_LSCO,2011_Chen_twomagnonFeSC}.

The peak energy of the magnetic excitation $\omega_\text{peak}$ takes into account the finite lifetime of the excitation by including the damping effect. We inspect it as a function of momentum, in Fig. \ref{fig:dispersion}(d), where it is possible to observe its energy dispersion as a function of transferred momentum for different dopings and directions. We found that the magnon dispersion is considerably softened for the Cr$8.5\%$ sample only, and equally softened in both directions, which contrasts with the Mn$8\%$ sample \citep{garcia_anisotropic_2019}.

To further investigate this effect, we can also analyze the trend of the bare frequency, $\omega_0$, and the damping coefficient $\gamma$, as shown in Figs. \ref{fig:dispersion}(e) and (f). The $\omega_0$ $\mathbf{q}$-dispersion is nearly sample independent, and we compare it with the results of a linear wave theory, plotted with parameters obtained from inelastic neutron scattering experiments of the parent compound \citep{harriger_nematic_2011}. A good agreement is observed in the probed momentum region. The shadowed area represents the L dependency of the linear wave model, based on a Hamiltonian with a weak out-of-plane interaction component. The X/Y anisotropy of the detwinned sample is evidenced in this model, with our twinned sample dispersion lying in between the curves for the $\Gamma$X and $\Gamma$Y directions.

Along $(\pi,0)$, $\gamma$ is nearly $\mathbf{q}$-independent for the BFA and Cr$3.5\%$ samples, and becomes $\mathbf{q}$-dependent for Cr$8.5\%$, increasing monotonically with $\mathbf{q}$. It suggests shorter magnon lifetimes, likely due to competing Néel-type fluctuations derived from Cr. Along $(\pi,\pi)$, the damping varies non monotonically for BFA and, in particular, for Cr$3.5\%$. For Cr$8.5\%$, where the bimagnon peak can be neglected, the damping increases monotonically with $\mathbf{q}$, indicating that a genuine momentum-dependent $\gamma$ emerges for CrBFA with increasing Cr content. Our results align with studies linking Cr doping in CrBFA to weaker orthorhombicity and more localized (yet still itinerant) magnetism above $x > 0.05$ \citep{clancy_high-resolution_2012} and theoretical modeling that increasing frustration may suppress the bimagnon peak in the FeSCs \cite{2011_Chen_twomagnonFeSC}.

\begin{figure}
    \centering
    \includegraphics[width=\linewidth]{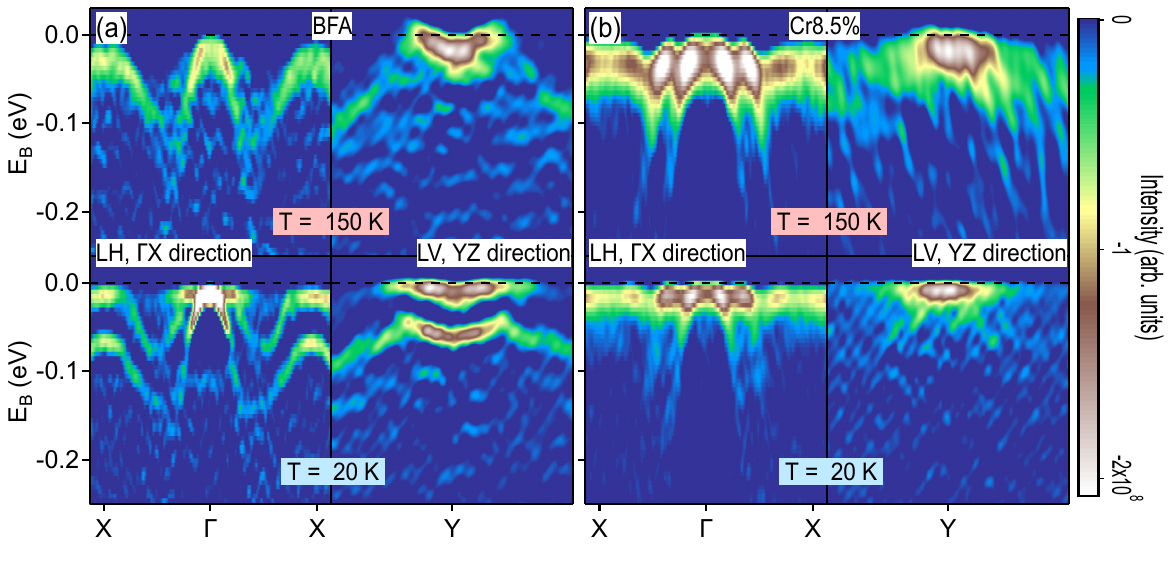}
     \caption{ARPES band maps second derivatives at paramagnetic and ordered phases for (a) BFA and (b) Cr$8.5\%$ samples. The transition temperatures are $T=134$ and $T=79$ K, respectively. The minimum of the second derivatives denotes the Lorentzian-shaped bands' maximum position.\label{fig:lowTARPES}}
\end{figure}

To understand the nature of the magnetically ordered state as a function of doping more comprehensively, we turn to ARPES measurements below and above $T_\text{N}$, as shown in Fig. \ref{fig:lowTARPES}. By inspecting the two different high-symmetry cut band maps, presented in the form of second derivatives of energy distribution curves (EDCs), we can access the band maximum points and see the electronic bands' dispersion shape. 

In the paramagnetic state, ARPES measurements of CrBFA reveal the increase in the hole pockets caused by Cr and the increasing electronic correlations, as characterized by the fractional scaling of the self-energy as a function of binding energy \citep{2024_cantarino_CrBFA}, a characteristic property of Hund's metals \citep{fernandes_iron_2022,georges_hund-metal_2024}.

At $T=20$ K, the BFA sample shows the typical $d_{xz/yz}$ orbital splitting from stripe-type itinerant order [Fig. \ref{fig:lowTARPES}(a)], reflecting the itinerant spin character in the AFM phase. The same reconstruction is present for Mn substitutions up to $8.5\%$ along both $\Gamma$X and YZ cuts \citep{cantarino_incoherent_2023}. In contrast, for $8.5\%$ Cr substitution [Fig. \ref{fig:lowTARPES}(b)], no such reconstruction is observed. Instead, the bands crossing the Fermi surface around X (mainly $d_{xy}$) become noticeably flatter, especially in the ordered state.

The absence of the electronic structure reconstruction in the case of CrBFA implies a more localized nature for the magnetic ground state at this doping, as also observed in previous experiments \citep{kobayashi_carrier_2016, clancy_high-resolution_2012} and is consistent with the weaker orthorhombicity discussed above, since the $d_{xz}/d_{yz}$ splitting is intrinsically tied to the structural and magnetic phase transitions. Motivated by these observations, we developed and solved a theoretical XXZ model of localized spins to calculate the effect of the doping disorder on the magnetic excitations' damping and dispersion, as detailed in the Supplemental Material \ref{appA_theory}.

Figure \ref{fig:theory} shows theoretical results (dashed magenta line). Panels (a) and (b) present the dynamical structure factor, including intensities and linewidths (damping) as a function of Cr content $x$. We adopted $J_1=18.75$ meV to reproduce the magnon bandwidth at $x=0$. The dashed white rectangles mark experimentally accessible regions that are shown in more detail in the insets, wherein the color maps are rescaled to ease comparison with the experiment. Panel (c) shows the structure factor at $\mathbf{q} = \bar{Y}$ versus energy for different $x$, while panel (d) presents, for comparison, the experimental magnon, after elastic and phonon subtraction, for the same $\mathbf{q}$ point. The finite width of the theoretical curve at $x=0$ is due to the finite-size resolution of the calculations. 

Our calculations explicitly include local fluctuations from Cr impurities and reproduce two key experimental observations: damping and softening of the excitations. The damping nearly doubles between $x = 0\%$ and $8.5\%$, matching the experimental trend. The suppression of the energy scale of the excitations (the softening), however, is stronger in experiments than in the theory, likely due to the missing itinerant–localized dual character of the spin excitations. This indicates that charge doping also contributes to magnon softening. Panels (e) and (f) show the experimental $\mathbf{q}$ dispersion maps for BFA and Cr$8.5\%$ along the same high-symmetry direction as in the panels (a) and (b) insets. Within the experimental  $\mathbf{q}$-range, theory and experiments agree with increasing spectral weight towards lower energies.

\begin{figure}
    \centering
    \includegraphics[width=\linewidth]{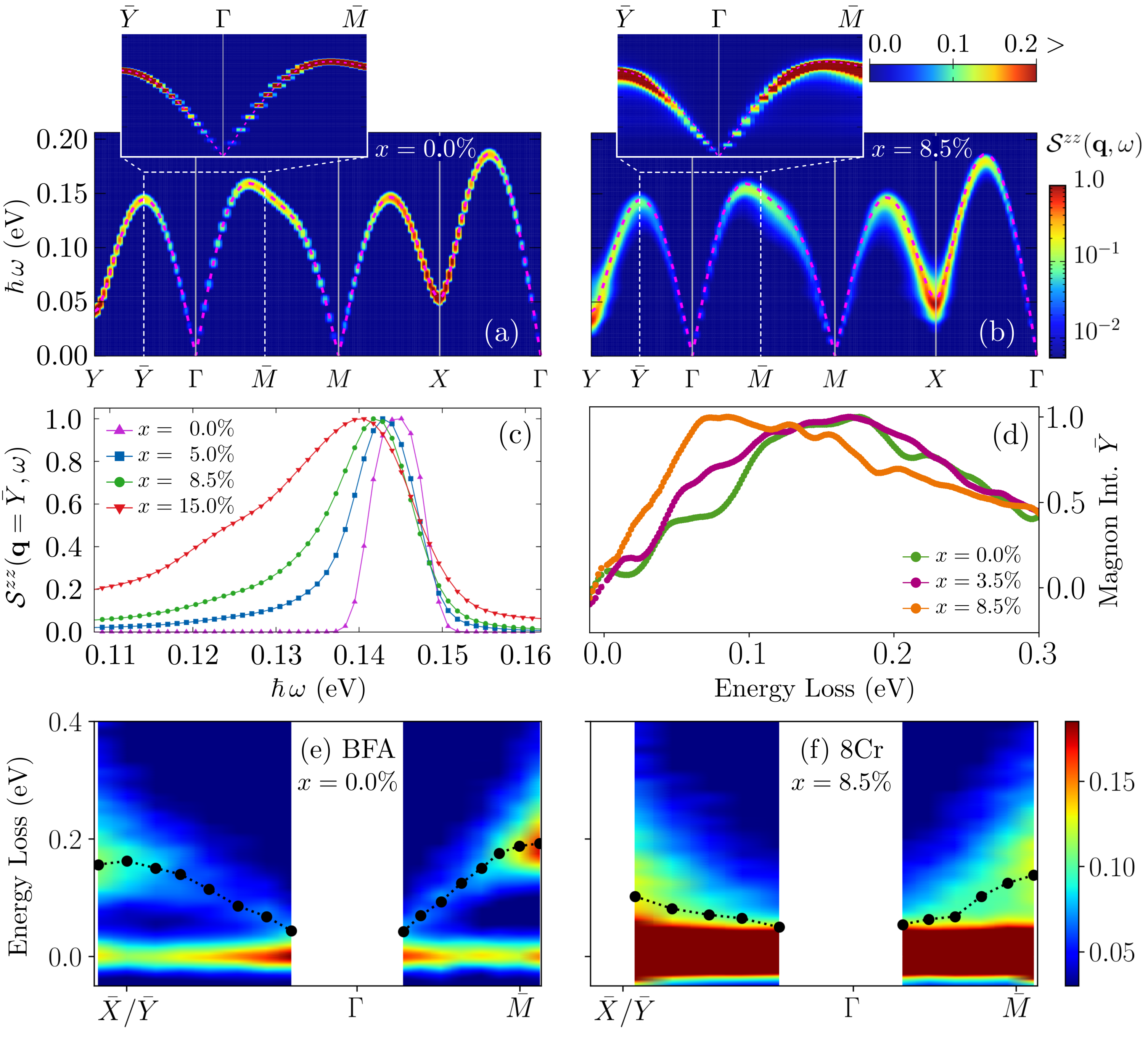}
    \caption{(a,b) Calculated dynamical structure factor of the $J_1$–$J_2$ XXZ model for different  doping levels ($x$). Insets show the RIXS probed region. The magenta dashed line marks the spin-wave dispersion for $x=0$. (c) Structure factor at $\mathbf{q}=\bar{Y}$ for different $x$. (d) Experimental magnon at $\mathbf{q}\approx (0.5,0)=\bar{X}/\bar{Y}$ for varying $x$ after elastic and phonon contributions subtraction. RIXS momentum maps for (e) BFA and (f) Cr$8.5\%$. Black dots are fitted $\omega_\text{peak}$, as in Fig.~\ref{fig:dispersion}(d).\label{fig:theory}}

\end{figure}

In addition to a good comparison with experiments, our calculations suggest that the magnon softening along the $(\pi,0)$ direction is related to a shift of Fe-derived spectral weight toward the $X/Y$ point with doping, as the magnetic excitations become more local, a $\mathbf{q}$-region inaccessible to RIXS. This theoretical insight may also help explain a key feature of the CrBFA and MnBFA phase diagrams: the absence of HTSC. Although Cr concentrates spectral weight near $(\pi,0)$/$(0,\pi)$, the associated excitations are highly incoherent (strongly damped), which is unfavorable for SC pairing \cite{WalberBrito2018,2018_Lee_pairing_Hund}.  

Other substitutional studies also indicate that local fluctuations from Cr moments can dominate over charge doping effects, as observed in Cr substitution of Ni-doped BFA \cite{Gong_LuoDai_2018_Cr-NiBFA_XRD,Zhang_LuoDai_2015_Cr-NiBFA_neutron,Gong_LuoDai_2022_Cr-NiBFA_ARPES} and P-substituted BFA \cite{Wenliang_LuoDai_2019_P-CrBFA_QCP,pelliciari_reciprocity_2019}. In CsFe$_2$As$_2$, disorder alone may even enhance correlations \cite{2023_Li_spinglass_CsFe2As2,crispino_paradigm_2023}.


\textit{Summary and conclusions:} Our RIXS experiments investigated the evolution of the magnon dispersion along high-symmetry directions as a function of Cr content in CrBFA. Combined with theoretical modeling, our results suggest a scenario where, across the CrBFA $x$ vs. $T$ phase diagram, disorder dominates the damping and partially explains the softening of the excitations. It reveals that charge doping mainly contributes to suppressing the excitations. This places disorder, instead of charge doping, as the primary effect in CrBFA, similarly to the case of Mn substitution, where ARPES observes minimal hole doping.

This is unexpected given the increasing electronic correlation, changing hybridizations, and electronic filling as observed by ARPES experiments of CrBFA \citep{2024_cantarino_CrBFA}. Our results show that disorder overrides these electronic effects in governing spin dynamics. We thus expect that our work may stimulate theoretical/experimental work on the effects of disorder on Hund's metals.  

\begin{acknowledgments}

We acknowledge the Diamond Light Source for RIXS beamtime at I21 under Proposal MM33194. We acknowledge support from the S\~ao Paulo Research Foundation (FAPESP), Grants Nos. 2019/05150-7 (M.R.C), 2019/25665-1 and 2024/13291-8 (F.A.G.), 2021/06629-4, 2022/15453-0, and 2023/06682-8 (E.C.A and R.M.P.T), 2017/10581-1 (K.R.P, P.G.P and C.A). E.C.A was supported by CNPq (Brazil), Grant No. 302823/2022-0, F.A.G by Grant No. 444081/2024-0. P.G.P. and C.A. acknowledge support from CNPq Grants No. 311783/2021-0 and 405408/2023-4. TS acknowledges financial support by the Swiss National Science Foundation through project no. 178867.

\end{acknowledgments}

\bibliography{2024_MCantarino_BFCA_references}
\pagebreak

\cleardoublepage

\makeatletter
\renewcommand \thesection{S\@arabic\c@section}
\renewcommand \thetable{S\@arabic\c@table}
\renewcommand \thefigure{S\@arabic\c@figure}
\setcounter{page}{1}
\makeatother

\widetext

\cleardoublepage

\begin{center}
\textbf{{\Large Supplemental Material}}
\end{center}

{
\twocolumngrid

\setcounter{equation}{0}

\section{Theoretical model for magnetic excitations}\label{appA_theory}

In this section, we provide details about the theoretical model and numerical methods adopted in our investigation of the spectrum of magnetic excitations when disorder on the exchange interactions is gradually introduced. We describe the classical Hamiltonian and our approach to disorder, then we discuss how the dynamical spin structure factor (DSSF) at a given temperature can be numerically estimated by combining Monte-Carlo (MC) with semiclassical molecular dynamics (SMD).

\subsection{Magnetic model and disorder due to impurities}

We consider the antiferromagnetic Heisenberg XXZ model \cite{Joyce1967,Manousakis1991,Auerbach1994,Parkinson2010} on a square lattice for a minimal description of the magnetic excitations. We work on the classical limit, $|\vec{S}|\to\infty\tsp$, and treat the localized magnetic moments (spins) in the lattice $\vec{S}(\mathbf{r}_{i})=S_{i}^{\mu}\tsp \widehat{e}_{\mu}$ as unit vectors (with $N$ being the total number of sites, $i=1,2,\dots,N-1,N$ and $\mu=x,y,z$). The anisotropic Hamiltonian with exchange interactions up to next-nearest-neighbors is given by $H=\nicefrac{1}{2}\sum_{\tsp i,j,\mu}J^{\mu}_{ij}\,S^{\mu}_{i}S^{\mu}_{j}$, where $J^{\tsp x,y}_{ij}=J_{ij}$ and $J^{\tsp z}_{ij}=\lambda\tsp J_{ij}$ with $J_{ij}\geq 0$ and $0 \leq \lambda < 1$ so that the Hamiltonian favors the ordering of spins in the $xy$ plane (we consider $\lambda=0.95$). In the clean system of Fe-sites, $J_{ij}=J_{1}$ for nearest-neighbors bonds and $J_{ij}=J_{2}$ for next-nearest-neighbors bonds, otherwise $J_{ij}=0$. The system exhibits Néel order for $J_{2}/J_{1}<1/2$ and stripe order for $J_{2}/J_{1}>1/2$. In our study, we gradually introduce disorder in the exchange interactions for next-nearest-neighbors bonds $\langle\langle i,j \rangle\rangle$ by means of doping with Cr magnetic impurities as shown in Table~\ref{FeCrTable}. The values presented in the latter were motivated by experimental observations and estimates based on inelastic neutron scattering data \cite{Ewings2008}. As the RIXS resolution is limited in energy and momentum, this simple model is sufficient to capture the spectrum key features. For a given concentration $x$ of impurities, we generate disorder configurations by randomly selecting $\lfloor\tsp x\,N \rfloor$ Fe-sites for substitution with Cr, each realization results in a list $D_{x}=\{\tsp i \tsp\}$ mapping the impurity distribution in the lattice.
\begin{table}[h]
    \centering
    \begin{tabular}{|c|c|c|}
        \hline
        $\;\;\left\langle\left\langle i,j \right\rangle\right\rangle\vphantom{\dfrac{x^{x}}{x}}$ bond type\;\; & $\;\;J_{2}/J_{1}\;\;$ \\
        \hline
        \hline
        Fe\;-\;Fe & 2 \\
        \hline
        Cr\;-\;Cr & 0 \\
        \hline
        Fe\;-\;Cr & 1 \\
        \hline
    \end{tabular}
    \caption{Coupling constants for exchange interactions between next-nearest neighbor sites in a square lattice, the ratio $J_{2}/J_{1}$ depends on the type of bond formed by Fe-sites and/or Cr-sites.\label{FeCrTable}}
\end{table}

\subsection{Numerical calculation of the DSSF}

The ensemble average of the energy-momentum resolved dynamical spin correlation function is the classical DSSF given by
\begin{equation}
    S^{\tsp\mu\nu}_{\textrm{\tsp\tiny C}}(\mathbf{q},\omega)=
    \dfrac{1}{2\pi N}\sum_{ij}\int_{-\infty}^{\tsp\infty}\!\!
    dt\;\langle S^{\mu}_{i}(t)S^{\nu}_{j}(0)\rangle\,
    e^{\img\omega t\tsp-\img\mathbf{q}\ccdot\mathbf{R}_{\hspace{0.2mm}ij}}\,.
    \label{DSSF}
\end{equation}
Here, $\mathbf{q}$ is a wave vector in reciprocal space, $\mu$ and $\nu$ are spin component indices, and $\mathbf{r}_{i}-\mathbf{r}_{j}=\mathbf{R}_{\tsp ij}$. 

We follow a semiclassical approach, as discussed in previous works \cite{Moessner1998,Conlon2009,Moessner2014,Shannon2018,Knolle2022}, to estimate this quantity using classical MC and SMD simulations: the former is employed to sample initial configurations of $N$ spins, $\lbrace S^{\mu}_{i}(t=0)\rbrace=\textrm{C}_{S}(0)$, and the latter to numerically integrate the equations of motion and obtain $\lbrace S^{\mu}_{i}(t\geq 0)\rbrace$. In our study, we consider a system of $N=L\times L$ spins, with $L=64$, at a fixed low-temperature value $T_{0}/J_{1}=10^{-2}$ and 4 different impurity concentrations $x\in\lbrace 0.00,0.05,0.10,0.15\rbrace$. In our numerical approach for the $x>0$ cases, we generate a set of 100 disorder configurations $\lbrace D_{x}\rbrace$ and for each one a set $\lbrace\textrm{C}_{S}(0)\rbrace$ with 100 spin configurations is sampled from MC simulations. The MC implementation is based on the Heat Bath \cite{Miyatake_1986} and the Microcanonical \cite{Creutz1987} local update algorithms, where each MC iteration consists of 9 microcanonical steps followed by 1 heat bath step (here, a step represents a random sweep over the entire lattice). Independent simulations are performed for each $D_{x}$ configuration, after 50000 MC iterations for equilibration, we start drawing the $\textrm{C}_{S}(0)$ samples at every 100. To avoid twinned samples (i.e., with horizontal or vertical stripe modulation), we always initialize the MC code with a spin configuration exhibiting horizontal modulation. In Fig.~\ref{SpinSamples}, two $\textrm{C}_{S}(0)$ samples are depicted within a small region of the $64 \times 64$ lattice. In (a), the concentration of impurities is small and the horizontal stripe ordering of the spins is barely altered. In (b), the concentration is high enough to induce a canting pattern that encompasses the whole lattice without disrupting the stripe order (moreover, small regions with higher local impurity concentration become common, enhancing the competition between Néel and stripe orders).

\setcounter{figure}{0}

\begin{figure}[t]
    \centering
    \includegraphics[width=0.95\linewidth]{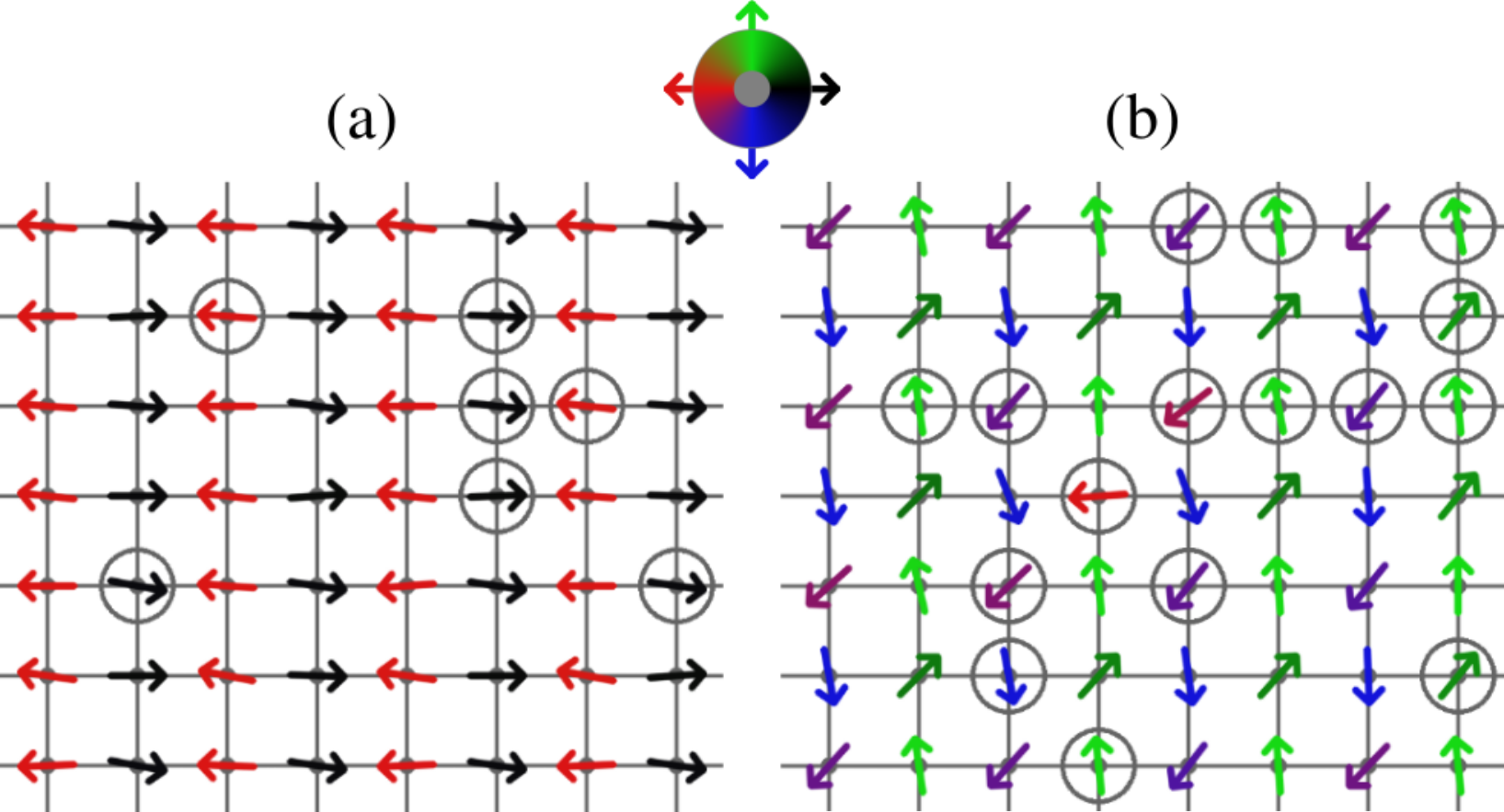}
    \caption{MC samples $\textrm{C}_{S}(0)$ for two cases: (a) $x=5\%$ and (b) $x=15\%$. Circles around spins represent sites where Fe were substituted by Cr. The transverse components $S^{z}_{i}$ are omitted, these are always very small compared to the inplane components $S^{\tsp x,y}_{i}$ which define the vectors in the figure colored accordingly to the inset 4-color disk. \label{SpinSamples}}
\end{figure}
 
In a separate numerical implementation, we use the data sets collected to estimate the classical DSSF from Eq.~\eqref{DSSF} by means of the average
\begin{equation}
    \begin{aligned}
        & \bar{S}^{\tsp\mu\nu}_{\textrm{\tsp\tiny C}}(\mathbf{q},\omega)=
        \dfrac{t_{\textrm{max}}}{2\tsp\pi}\left[\left\langle
        S^{\mu}_{\mathbf{q}}(\omega)\tsp S^{\nu}_{-\mathbf{q}}(-\omega)
        \phfrac\right\rangle_{\!\textrm{C}_{S}(0)}\tsp\right]_{D_{x}}\!,
        \label{nDSSF}
        \\
        & S^{\mu}_{\mathbf{q}}(\omega)=
        \dfrac{1}{\sqrt{N^{\vphantom{X}} N_{t}}}\sum_{i,n}S^{\mu}_{i}(t_{n})
        \exp(\img\omega\tsp t_{n}\tsp-\img\mathbf{q}
        \cdot\mathbf{r}_{i})\;.
    \end{aligned}
\end{equation}
Here, $\langle\tsp\dots\rangle$ represents the ensemble average (i.e., the MC estimation based on the samples $\lbrace\textrm{C}_{S}(0)\rbrace\tsp$) whereas $[\tsp\dots]_{D_{x}}$ denotes the average over disorder configurations (absent for the clean case $x=0$). Each contribution to the combined average in Eq.~\eqref{nDSSF} involves the Fourier transform in space and time $S^{\mu}_{\mathbf{q}}(\omega)$ of the temporal series $\lbrace S^{\mu}_{i}(t_{n})\,|\,t_{n}=n\tsp\delta t\tsp,\,n=0,1,\dots,N_{t}\rbrace$ for some spin configuration, where $N_{t}$ is the number of time steps within a finite time-interval with $t_{\textrm{max}}=N_{t}\tsp\delta t$ representing the final evolution time reached under a fixed temporal discretization $\delta t$. The evolution equation can be derived from Heisenberg's equation of motion for spin operators $d\tsp S^{\mu}_{i}(t)/dt=\textrm{i}\left[H,S^{\mu}_{i}(t)\right]/\hbar$ in the classical limit \cite{Auerbach1994}, it describes the precession dynamics of each spin about the local field $h^{\mu}_{i}=\sum_{\tsp j}J_{ij}^{\mu}\tsp S_{j}^{\mu}$ produced by its neighbors: 
\begin{equation}
    \dfrac{d\vec{S}_{i}(t)}{dt}=
    \vec{h}_{i}(t)\times\vec{S}_{i}(t)\;.
    \label{SpinDyn}
\end{equation}
In SMD simulations, we employ the fourth-order Runge-Kutta (RK4) algorithm \cite{Dormand1980} combined with the energy correction (EC) algorithm from Ref. \cite{Garanin2021} to obtain $\lbrace S^{\mu}_{i}(t_{n})\rbrace$ given some $\textrm{C}_{S}(0)$ sample as initial condition. We use the Fast Fourier Transform (FFT) \cite{Frigo2005} algorithm to calculate $S^{\mu}_{\mathbf{q}}(\omega_{k})$, with $\omega_{k}=k\tsp\delta\omega$ being discrete frequencies ($k=-N_{t}/2,-N_{t}/2+1,\dots N_{t}/2-1$). In order to obtain a suitable frequency resolution $\delta\omega=2\tsp\omega_{\textrm{max}}/N_{t}$ over the entire frequency range while keeping $\delta t$ small for the RK4\plus EC algorithm, we use $\omega_{\textrm{max}}=120$ and $N_{t}=4000$ which gives $\delta\omega=0.06$. The latter implies that $\delta t=\pi/\omega_{\textrm{max}}\approx 0.026$ and $t_{\textrm{max}}\approx 104.7$. With these parameters, our SMD code preserves very well the system energy $H(t)$ after each time step, with the final relative difference $\delta_{0} H=[H(t_{\textrm{max}})-H(0)]/H(0)$ of the order of $10^{-12}$ (without EC, the energy drifts away from $H(0)$ linearly and $\delta_{0} H$ is of order of $10^{-5}$).

We mention that the Fourier transform in time is applied to the input data $w(t_{n})\tsp S^{\mu}_{\mathbf{q}}(t_{n})$, where the leading factor is a window function (usually a Gaussian envelope) that avoids numerical artifacts due to the FFT algorithm assumed periodicity. We use a custom window function that balances resolution and spectral leakage. It is designed to have a raised-cosine response with a very narrow transition from passband to stopband, and is approximated using a Kaiser FIR (finite-impulse-response) window/filter with appropriate parameters.

In order to compare our numerical results for the classical DSSF of Eq.~\eqref{nDSSF} with experiments and theoretical approaches such as LSWT (linear spin wave theory), we follow the Refs. \cite{Paddison2019,Smith2022,Knolle2022} and rescale $\bar{S}^{\tsp\mu\nu}_{\textrm{\tsp\tiny C}}(\mathbf{q},\omega)$ by the Bose-Einstein distribution $1/[1-\exp(-\beta E)]$ multiplied by $\beta E$, where $E=\hbar\omega$ is the excitation energy and $\beta=1/(k_{B}T)$ the inverse temperature. With this correction factor, we obtain the quantum-equivalent DSSF as (setting $\hbar=1$)
\begin{equation}
    \mathcal{S}^{\tsp\mu\nu}(\mathbf{q},\omega)=
    \dfrac{\beta\tsp\omega}{1-\exp(-\beta\tsp\omega)}\,
    \bar{S}^{\tsp\mu\nu}_{\textrm{\tsp\tiny C}}(\mathbf{q},\omega)\;.
    \label{nQDSSF}
\end{equation}
The intensity plots (in arbitrary units) of the DSSF presented in the main text correspond to our results obtained from Eq.~\eqref{nQDSSF} for the longitudinal contribution $\mathcal{S}^{zz}(\mathbf{q},\omega)$. To enhance the visibility of spectral features across the energy-momentum regions of interest, we normalize the spectrum intensity and impose a high intensity cutoff at $20\%$ (when the region is the whole Brillouin zone, we instead impose a fixed low intensity cutoff prior to applying a logarithmic scale). For comparison with experiments, we use $J_{1}=18.75$ meV, which reproduces the $x=0$ magnon bandwidth, and it is compatible with inelastic neutron scattering results \cite{Ewings2008}. Additionally, we superimpose a dashed line indicating the LSWT energy dispersion that we derived for the Heisenberg XXZ model considered in this study. In Fig. \ref{DSSF_TwoPaths}, we show some results in two energy-momentum regions for two doping levels.

\begin{figure}[h!]
    \centering
    \includegraphics[width=1.0\linewidth]{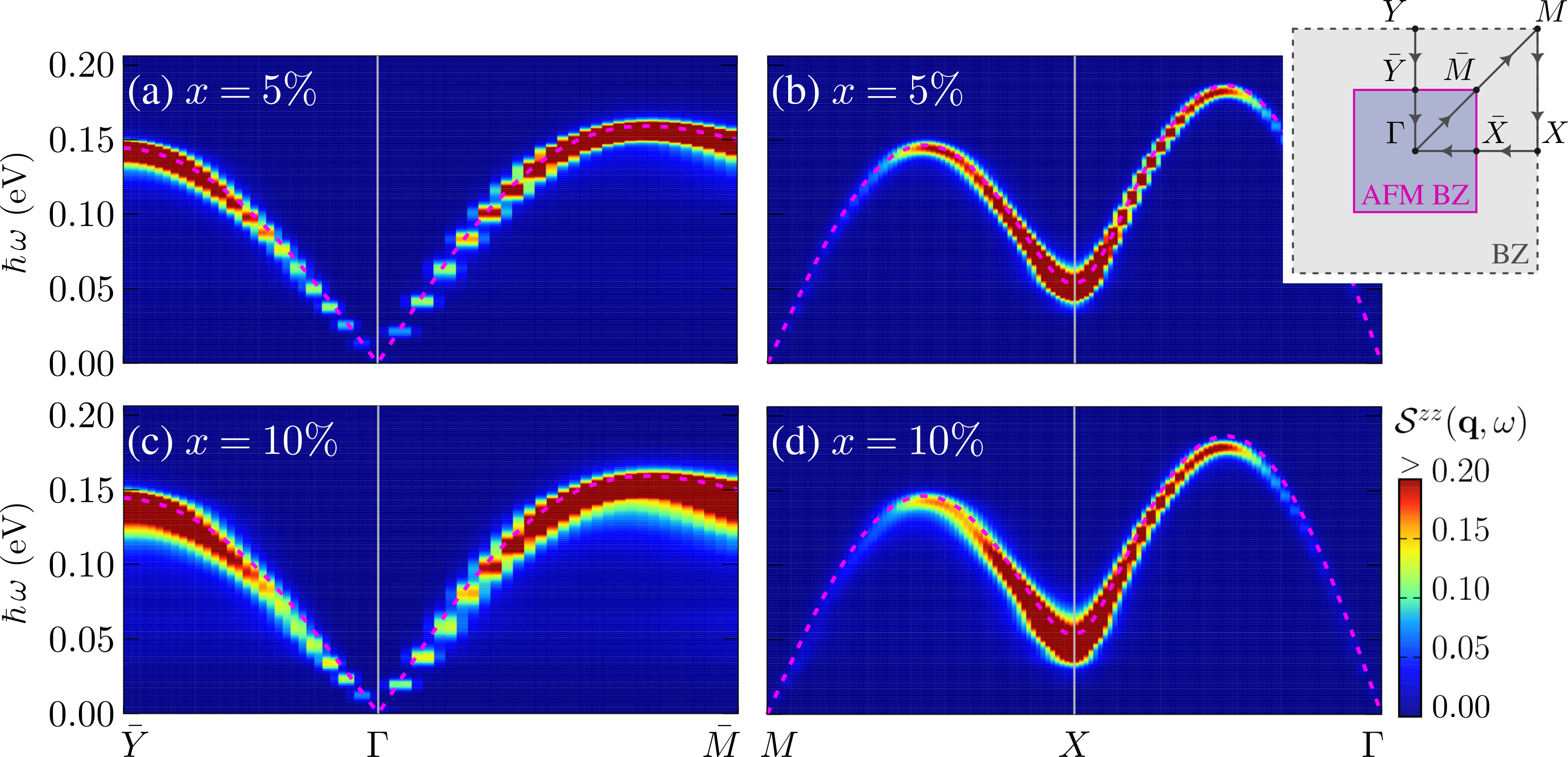}
    \caption{Calculated dynamical structure factor of the $J_1$–$J_2$ XXZ model for two doping levels ($x$). (a) and (c) show the results within the RIXS probed region, while (b) and (d) focus on a region (inaccessible to RIXS) where most of the spectral weight is concentrated. The magenta dashed line marks the spin-wave dispersion at $x=0$. \label{DSSF_TwoPaths}}
\end{figure}
}

\widetext
\cleardoublepage

\section{Experimental Methods}\label{appB_exp}

\textit{Sample characterization}:
The samples were characterized by resistivity, specific heat, x-ray diffraction (XRD), and energy-dispersive x-ray spectroscopy (EDS) for their transition temperatures $T_{\text{N}}$, lattice parameters, and chemical content, respectively. The results were compared to composition vs. $T$ phase diagrams in literature \citep{sefat_absence_2009,marty_competing_2011,clancy_high-resolution_2012} to benchmark the values of $x$.

\textit{RIXS experimental details}:
The samples were glued onto Cu sample holders using silver epoxy and cleaved in a vacuum of $1\times10^{-8}$ Torr, using Al posts, at room temperature. The samples were then transferred to the measurement chamber, with a vacuum of $4\times10^{-10}$ mbar. Experiments were performed in the magnetically ordered phase of BFA, with $T = 15$ K. The scattering angle was fixed at $154^\circ$, for the maximum in-plane momentum transfer, and the sample angle was rotated so the momentum projection could be changed. The maximum momentum transfer was achieved for a grazing incidence geometry of $13^\circ$. 

Linear horizontal (LH) and vertical (LV) polarized $X$-rays were used to probe the absorption spectrum (XAS) of each sample, and an energy-dependent RIXS map was done to study the incident energy dependence of the observed features. The energy of the absorption spectra maximum at $709.5$ eV and LH-polarized light with grazing incidence were chosen for the momentum-dependent studies. The RIXS experiments were carried out along the crystallographic directions $[110]$ and $[100]$. This corresponds, respectively, to iron sub-lattice first neighbors' and second neighbors' directions, also denoted as $(\pi,0)$ and $(\pi,\pi)$, or high-symmetry directions $\Gamma X$ and $\Gamma M$ of the reciprocal space, consistent with our previous ARPES results convention. For each RIXS spectrum, a carbon tape measurement was performed to determine precisely the energy resolution and zero energy transfer position. 

\textit{ARPES experimental details}:
ARPES data were measured at the Bloch beamline of the Max IV synchrotron in Lund, Sweden. The total energy resolution was $\sim 10$ meV for incident photon energy of $76$ eV, and angular resolution of $0.1^\circ$. The LH polarized $\Gamma$X direction band map presented in this work is a high-statistic band measurement. The LV polarized YZ direction band map was reconstructed from Fermi Surface maps measurements. We present second derivative band maps for discussing band position, size and shape.

\begin{figure}[h]
    \centering
    \includegraphics[width=0.75\linewidth]{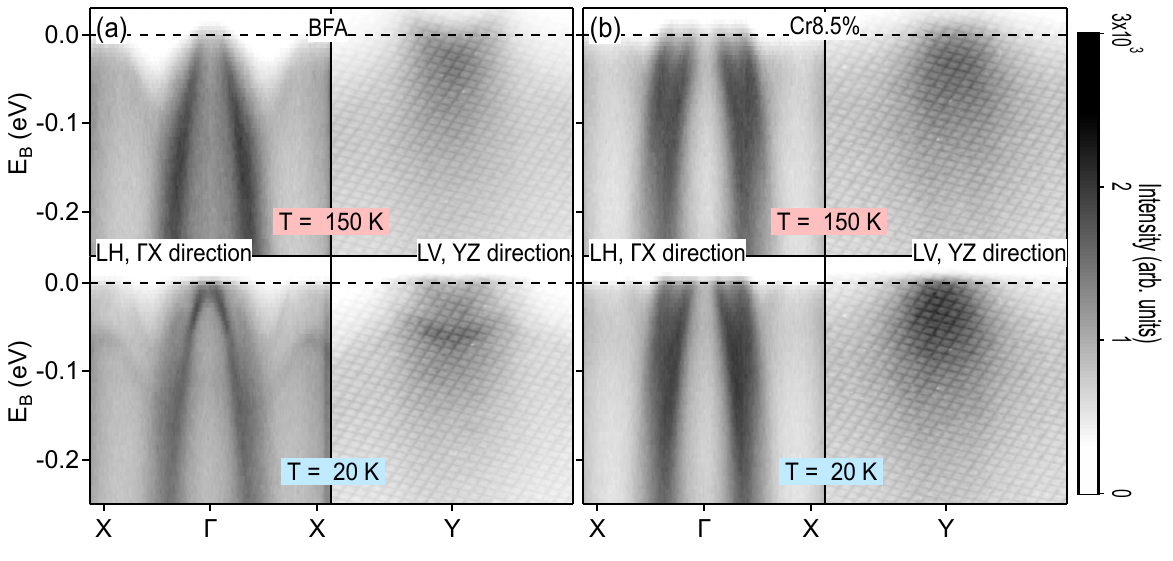}
     \caption{ARPES band maps at paramagnetic and ordered phases for (a) BFA and (b) Cr$8.5\%$ samples.\label{fig:rawARPES}}
\end{figure}

Fig. \ref{fig:rawARPES} shows the original intensity band map of the second derivative data presented in Fig. \ref{fig:lowTARPES}. For the BFA sample, it is evident that there is a band splitting as the sample is cooled, associated with the nematic orbital degeneracy lift. 

\subsection{Low-energy RIXS background subtraction and fitting}\label{appRIXS_fit}

All data presented in this study were normalized by the maximum fluorescence intensity before its subtraction, allowing a proper comparison of the quasi-elastic and magnon peak intensity as a function of momentum and Cr concentration.

\begin{figure}[h]
    \centering
    \includegraphics[width=0.5\linewidth]{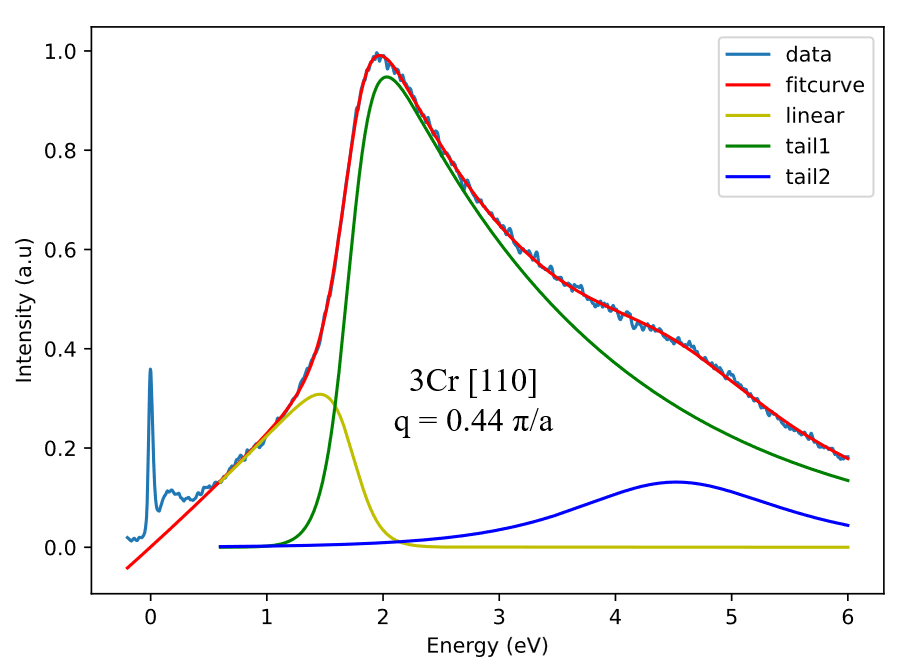}
     \caption{Example of a background fitting. The full spectrum of sample Cr$3.5\%$ measured for $[110]$ crystal orientation and the angle corresponding to momentum transfer of $0.44 \pi/a$ is shown as the light blue solid line. The resulting fitting curve is shown as the red line, and its three different contributions are yellow, green, and blue lines. The \label{fig:Fluo_fit}}
\end{figure}

The fluorescence background was fitted to a phenomenological model introduced by Yang \textit{et al.} \cite{2009_Yang_weakcor} as implemented already for the FeTe and BFA compounds \cite{2010_Hancock_FeTe,zhou_persistent_2013}. The model is illustrated in \ref{fig:Fluo_fit} and it separates the background into three regions with two crossover functions between them. The regions correspond to a linear behavior close to energy transfer $E \approx -1$ eV, shown as a yellow line, one exponential curve in the region $-4.5 \lessapprox E \lessapprox -2$ eV, shown as a green line, and another exponential curve in the region $-6 \lessapprox E \lessapprox -4.5$ eV, represented by a navy blue line. The complete model is depicted with the red line, which is a sum of the other three lines.

This model is described in equation \ref{eq:fluo_fullbckg}, where the three curves are weighted by coefficients $\alpha$, $\beta$, and $\gamma$. In this model, the frequency $\omega$ corresponds to the transferred energy, which is defined as negative by $E = E_d - E_i$, where $E_d$ is the detected photon energy and $E_i= \mu h$ is the incident photon energy. The parameters $a$, $b$, and $c$ determine the exponential shape.

\begin{equation}
    I_{fluo}=\alpha \exp(-a\omega)\omega(1-g_{\Gamma_1})+\beta \exp(b\omega)g_{\Gamma_1}
    +\gamma \exp(c\omega)g_{\Gamma_2}
    \label{eq:fluo_fullbckg}
\end{equation}

The function that is responsible for the crossover between the curves is described in equation \ref{eq:crossover}, where $\omega_{1,2}$ is the crossover position in transferred energy $E$ and $\Gamma_{1,2}$ determines the width of the crossover region.

\begin{equation}
    g_{\Gamma_{1,2}}=(\exp(-(\omega-\omega_{1.2})/\Gamma_{1,2})+1)^{-1}
    \label{eq:crossover}
\end{equation}

\begin{figure*}[h]
    \centering
    \includegraphics[width=\linewidth]{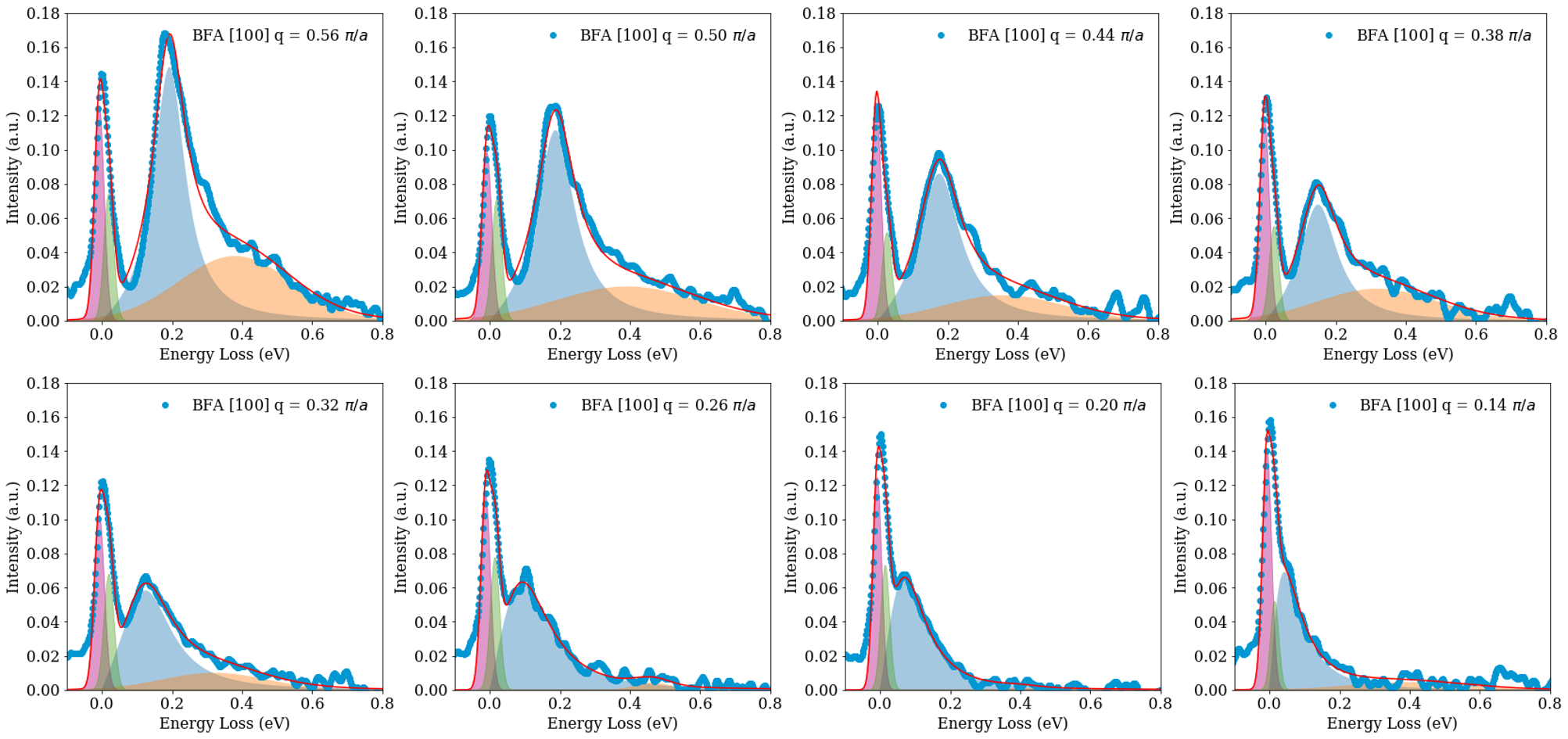}
     \caption{Momentum dependent magnon and quasi-elastic fittings for the BFA sample and $[100]$ direction. The data after background subtraction is shown as blue dots. The total fitting is the red line and the shaded regions are the fitting contributions, including the elastic line, phonon, main magnon, and secondary magnon peak, shown as pink, green, blue, and orange areas, respectively.\label{fig:BFA100}}
\end{figure*}

Using this function to subtract the fluorescence contribution allows us to have a longer zero line in energy loss for the resulting quasi-elastic plus magnon spectra, improving the fitting quality evaluation, such as residuals. To fit the quasi-elastic plus magnon resulting spectra, we use a model composed of two resolution-defined width Gaussians for the elastic line and phonon peak, one damped harmonic oscillator for the magnon peak, and a Gaussian for the secondary magnon peak.

\begin{figure*}[h]
    \centering
    \includegraphics[width=\linewidth]{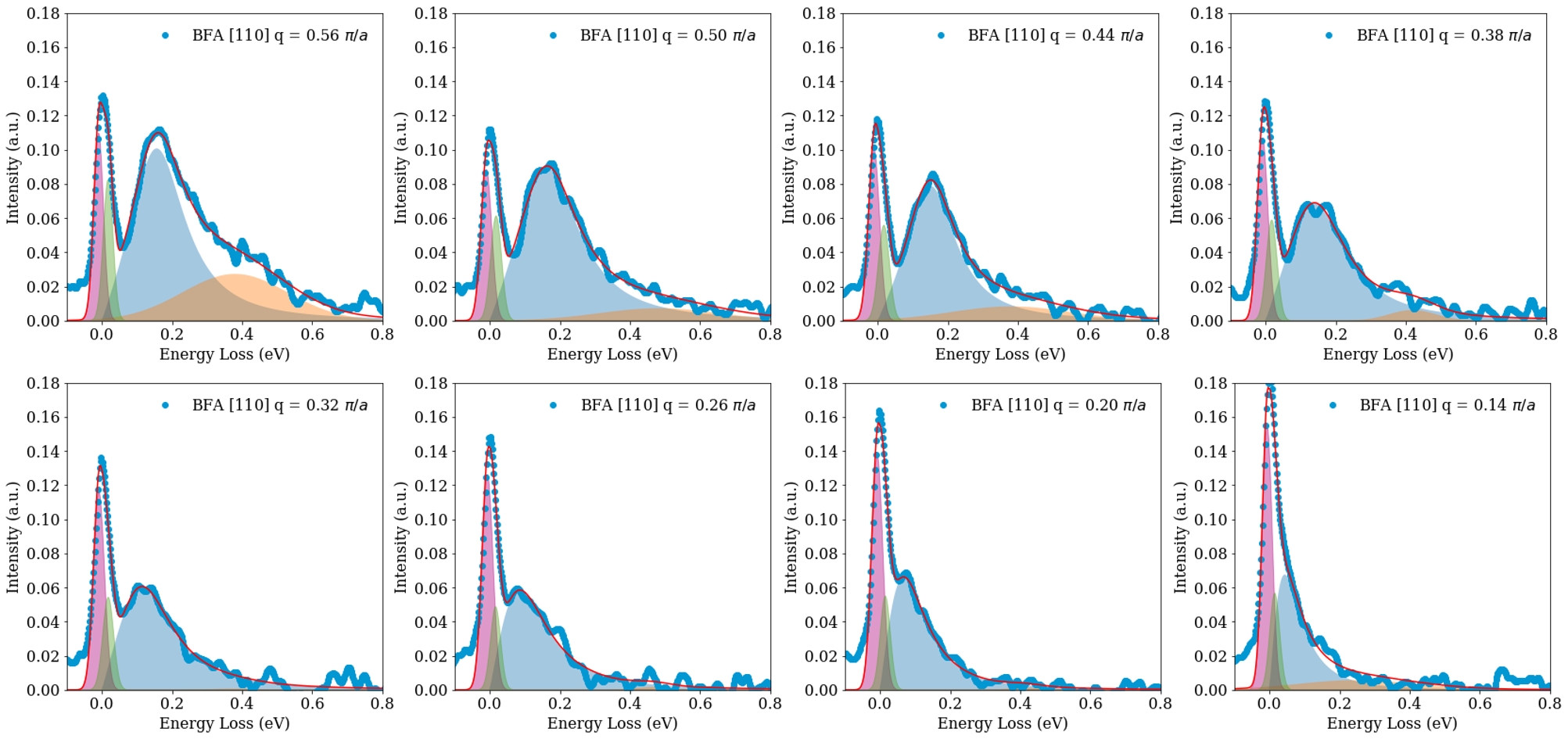}
     \caption{Momentum dependent magnon and quasi-elastic fittings for the BFA sample and $[110]$ direction. The data after background subtraction is shown as blue dots. The total fitting is the red line and the shaded regions are the fitting contributions, including the elastic line, phonon, main magnon, and secondary magnon peak, shown as pink, green, blue, and orange areas, respectively. \label{fig:BFA110}}
\end{figure*}

\begin{figure*}[h]
    \centering
    \includegraphics[width=\linewidth]{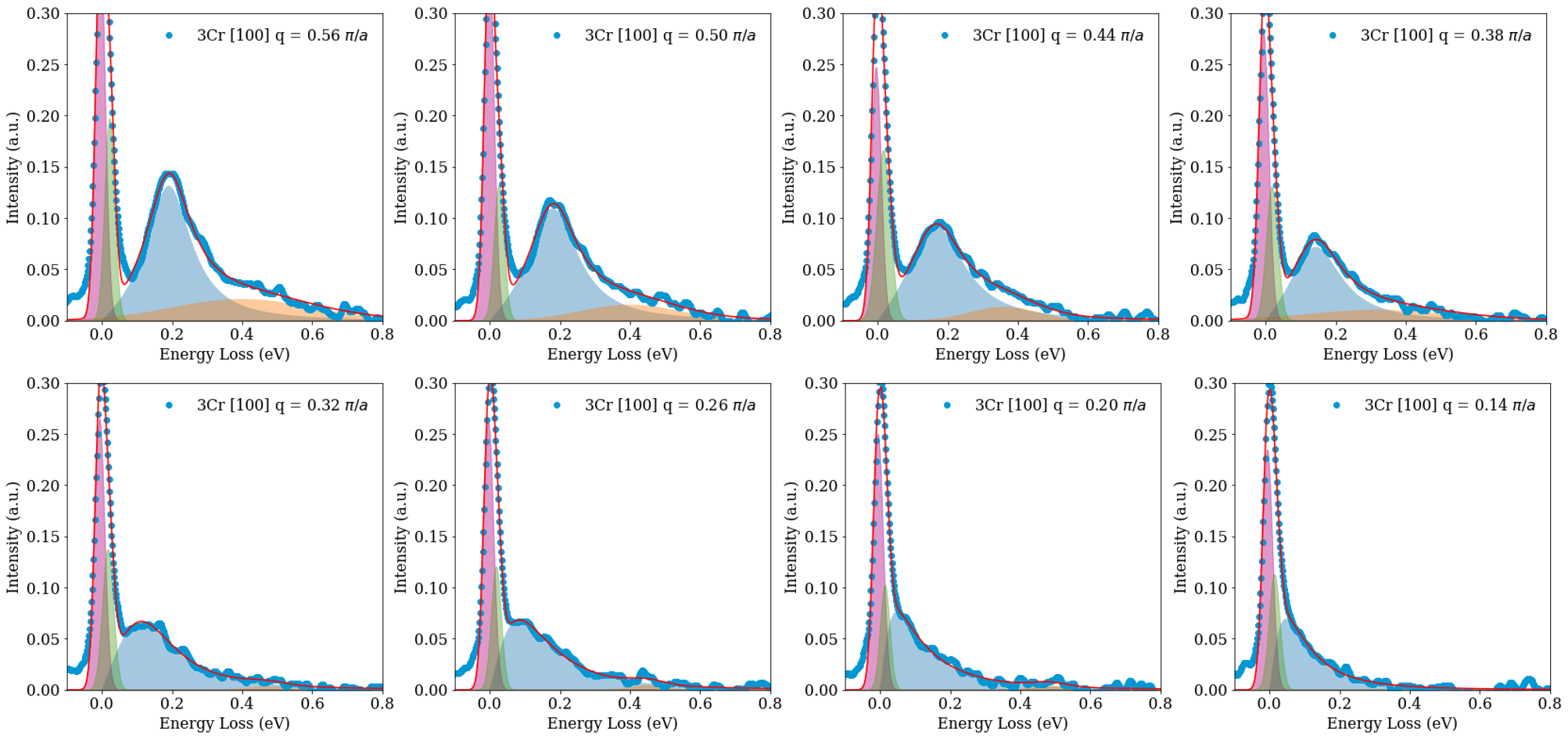}
     \caption{Momentum dependent magnon and quasi-elastic fittings for the Cr$3.5\%$ sample and $[100]$ direction. The data after background subtraction is shown as blue dots. The total fitting is the red line and the shaded regions are the fitting contributions, including the elastic line, phonon, main magnon, and secondary magnon peak, shown as pink, green, blue, and orange areas, respectively. \label{fig:3Cr100}}
\end{figure*}

\begin{figure*}[h]
    \centering
    \includegraphics[width=\linewidth]{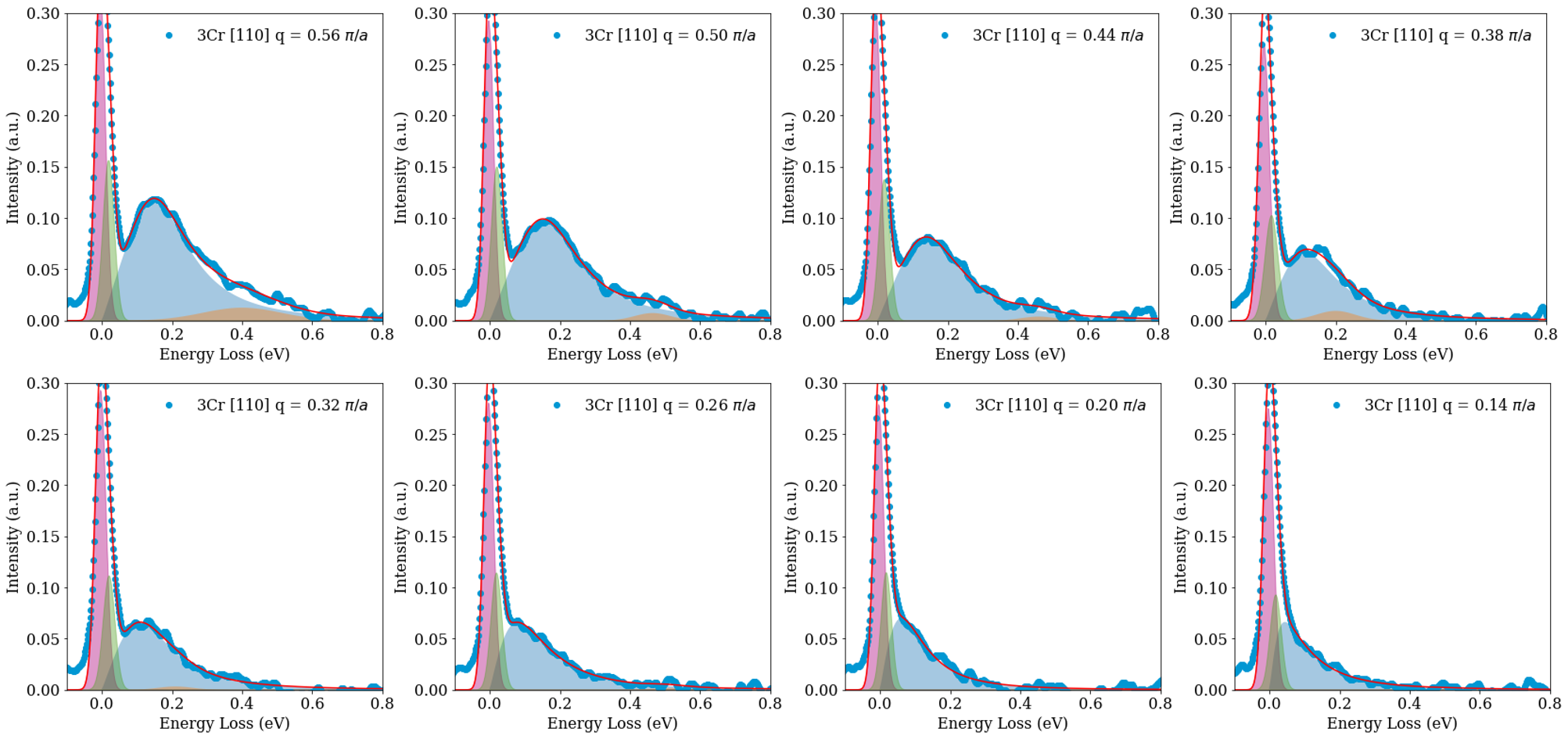}
     \caption{Momentum dependent magnon and quasi-elastic fittings for the Cr$3.5\%$ sample and $[110]$ direction. The data after background subtraction is shown as blue dots. The total fitting is the red line and the shaded regions are the fitting contributions, including the elastic line, phonon, main magnon, and secondary magnon peak, shown as pink, green, blue, and orange areas, respectively.  \label{fig:3Cr110}}
\end{figure*}

\begin{figure*}[h]
    \centering
    \includegraphics[width=\linewidth]{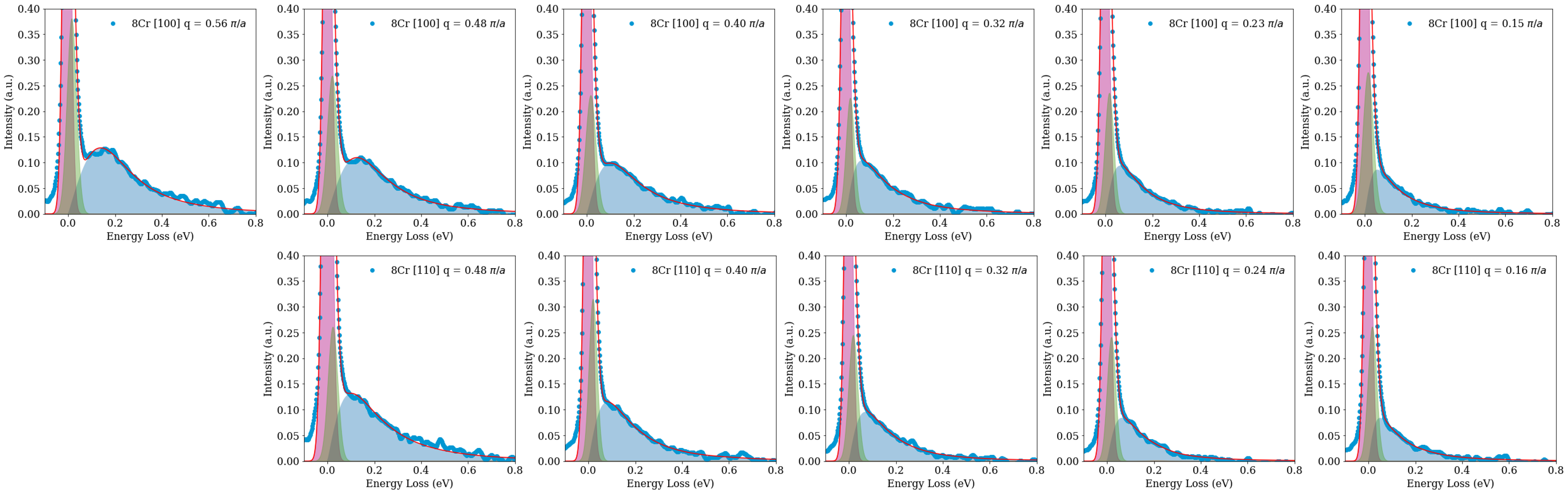}
     \caption{Momentum dependent magnon and quasi-elastic fittings for the Cr$3.5\%$ sample for both $[100]$ and $[110]$ directions. The data after background subtraction is shown as blue dots. The total fitting is the red line and the shaded regions are the fitting contributions, including the elastic line, phonon, main magnon, and secondary magnon peak, shown as pink, green, blue, and orange areas, respectively.  \label{fig:8Crfit}}
\end{figure*}

In Figs. \ref{fig:BFA100}-\ref{fig:8Crfit} is possible to see all the resulting quasi-elastic spectra and their respective fittings, with all included contributions. The total fitting is the red line and the shaded regions are the fitting contributions, including the elastic line, phonon, main magnon, and secondary magnon peak, shown as pink, green, blue, and orange areas, respectively. We can observe the quality of the fitting when the background is properly subtracted, making the need for an additional magnetic peak evident. The result phonon is momentum and sample independent, with energy loss of 19(2) meV.

\begin{figure*}[h]
    \centering
    \includegraphics[width=\linewidth]{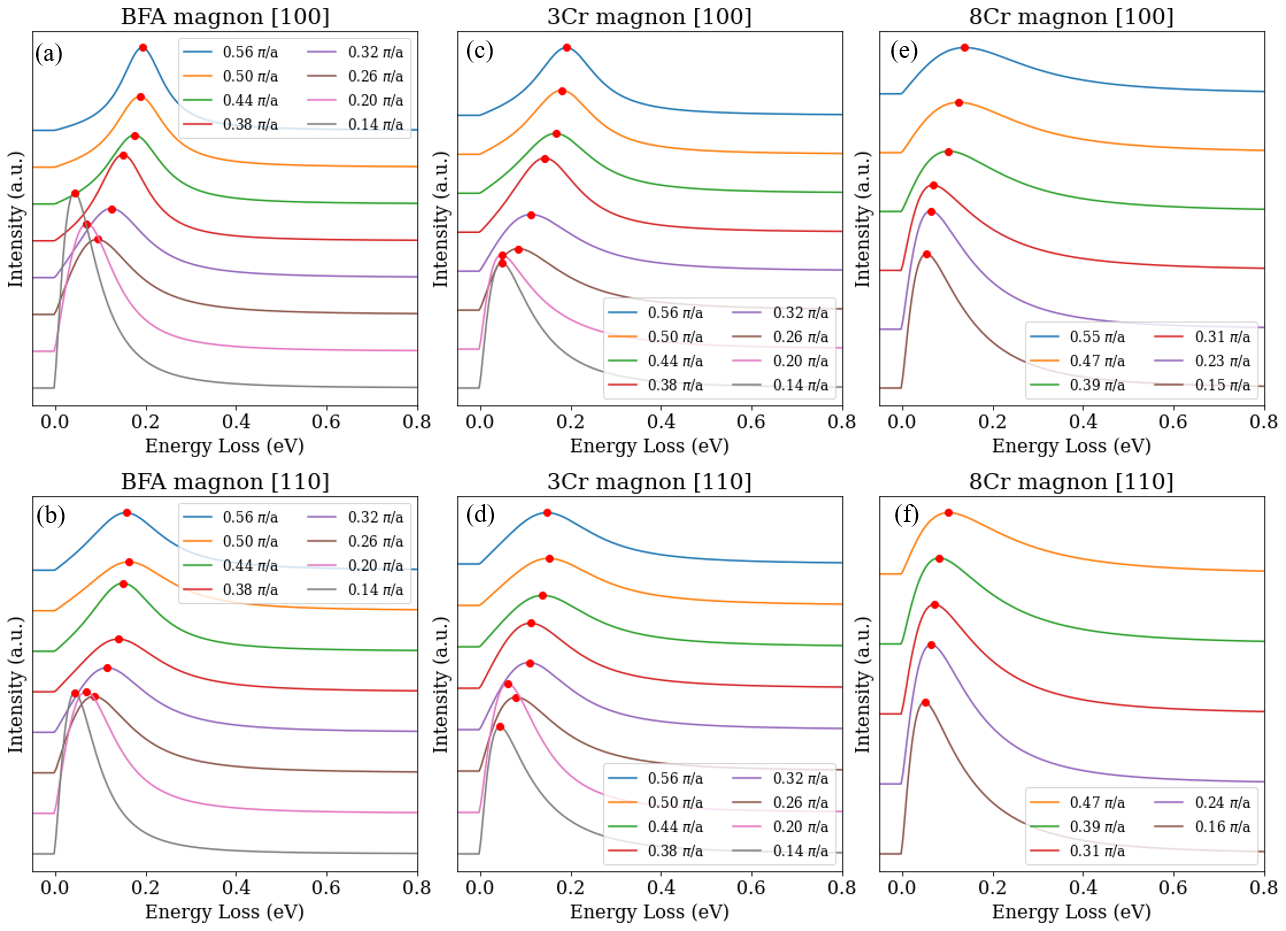}
     \caption{Main magnon peak fitted function, normalized by the $I_0$ fitted intensity, as a function of momentum for BFA sample for (a) $[100]$ and (b) $[110]$ directions, for Cr$3.5\%$ sample (c) $[100]$ and (d) $[110]$ directions and for Cr$8.5\%$ sample for (e) $[100]$ and (f) $[110]$ directions. \label{fig:magnonsfit}}
\end{figure*}

Figure \ref{fig:magnonsfit} shows the trend of the fitted main magnon peak, normalized by the fitted intensity $I_0$. An offset is included to make the trend more evident, and the peak energy of the magnetic excitation $\omega_\text{peak}$ is marked by a red dot for each momentum. We can observe the dispersion and shape of the fitted magnon as a function of momentum, direction, and doping.

\begin{figure*}[h]
    \centering
    \includegraphics[width=0.6\linewidth]{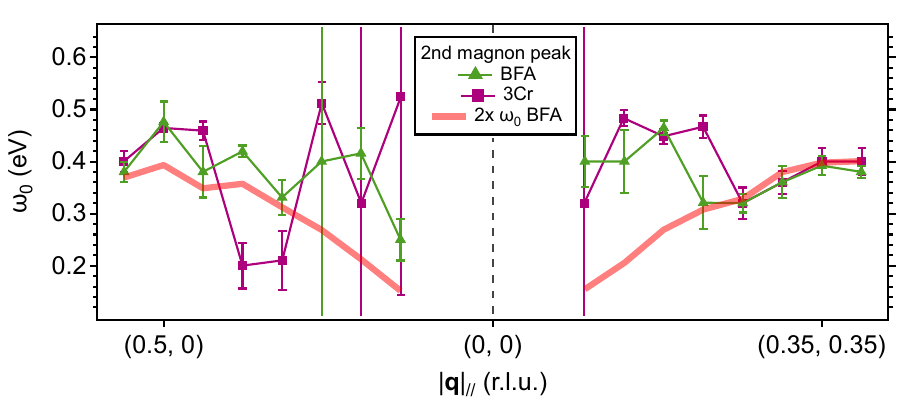}
     \caption{Bare frequency $\omega0$ of the second magnon peak versus momentum for BFA and Cr$3.5\%$ samples. The red line shows twice the $\omega_{0}$ of the main peak for the BFA sample. \label{fig:2ndmag}}
\end{figure*}

\begin{figure}
    \centering
    \includegraphics[width=0.5\linewidth]{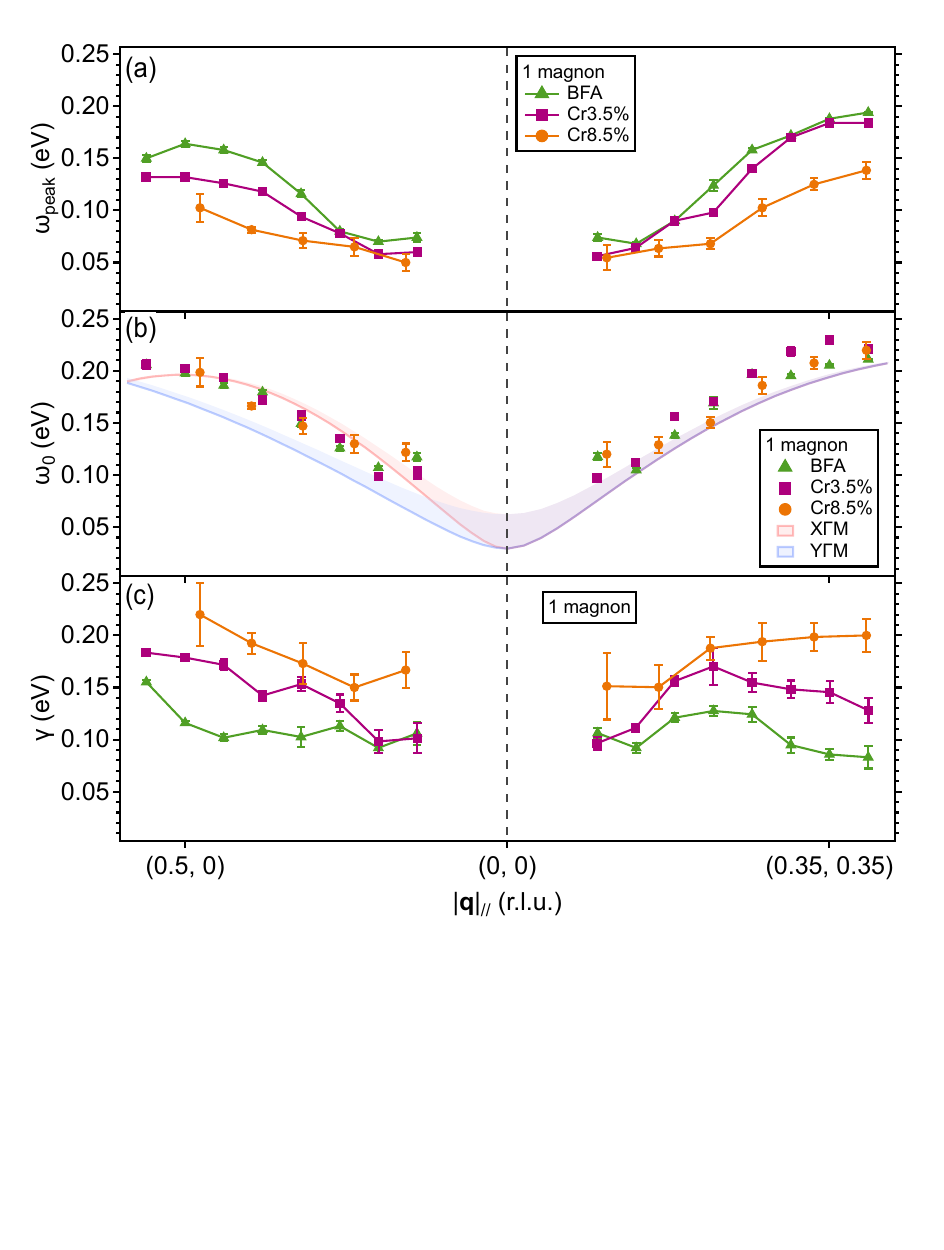}
    \caption{Results from fitting one magnon. (a) Peak energy of the magnetic excitation $\omega_\text{peak}$, (b) bare frequency $\omega_0$, and (c) damping coefficient $\gamma$ as a function of momentum, for different samples and high-symmetry directions.\label{fig:1mag_trend}}
\end{figure}

Figure \ref{fig:2ndmag} shows the dispersion of the bare frequency $\omega0$ associated with the second magnon peak, with clear momentum dependence where its contribution is significant (orange peaks in Figs. \ref{fig:BFA100}-\ref{fig:8Crfit}). The Cr$8.5\%$ data was omitted due to negligible relevance. At higher $\mathbf{q}$, the dispersion of this secondary peak follows approximately twice the main magnon energy (red line), supporting its identification as a bimagnon excitation.

Figure \ref{fig:1mag_trend} shows the trends of the peak energy of the magnetic excitation $\omega_\text{peak}$, bare frequency $\omega_0$, and damping coefficient $\gamma$ for all the samples when the spectra are fitted using a single-magnon model. Under this constraint, the damping regime of Cr$3.5\%$ is closer to that of Cr$8.5\%$, which is qualitatively different from what is observed from the raw data. Another key difference is that the softening in the Cr$3.5\%$ sample becomes more anisotropic, resembling the behavior previously reported for Mn$8\%$.

\FloatBarrier

\cleardoublepage

\twocolumngrid

\end{document}